\documentclass[twoside,english,final,3p,times,twocolumn,utf-8]{elsarticle}
\usepackage[T1]{fontenc}
\usepackage[latin9]{inputenc}
\usepackage{array}
\usepackage{pifont}
\usepackage{units}
\usepackage{amsbsy}
\usepackage{amstext}
\usepackage{amssymb}
\usepackage{graphicx}

\makeatletter

\providecommand{\tabularnewline}{\\}

\@ifundefined{date}{}{\date{}}
\usepackage{blindtext, graphicx, amsmath, algorithm, algpseudocode, pifont, algcompatible, comment, layout, amsthm, amssymb}
\usepackage{enumitem}   
\usepackage{eso-pic}
\usepackage{float}
\usepackage{lineno}
\modulolinenumbers[5]
\usepackage{wrapfig}
\usepackage{tikz}
\usetikzlibrary{fit}

\usepackage{multicol}
\usepackage[algo2e,linesnumbered,ruled,vlined]{algorithm2e} 
\usepackage{hyperref} 
\definecolor{hGreen}{HTML}{21897E}
\definecolor{dnBlue}{HTML}{2F5597}
\definecolor{nBlue}{HTML}{D7E3FC}
\definecolor{nPink}{HTML}{F20089}
\definecolor{nViolet}{HTML}{7209B7}
\usepackage{colortbl}
\usepackage{tabularx}
\hypersetup{ colorlinks=true, linkcolor=black, filecolor=black, urlcolor=cyan, }

\SetCommentSty{mycommfont}
\SetProcNameSty{textsc}

\journal{ICT Express}

\usepackage{dsfont}
\usepackage{epstopdf} 
\usepackage[tikz]{mdframed}
\usepackage{xcolor}
\mdfsetup{
middlelinecolor=black,
middlelinewidth=1pt,
backgroundcolor=red!10,
roundcorner=10pt}

\makeatother

\usepackage{babel}
\begin{document}

\begin{frontmatter}{}

\title{Nearest neighbor Methods and their Applications in Design of 5G \&
Beyond Wireless Networks}

\vspace{-1in}

\author{S.~A.~R.~Zaidi~\corref{cor1}}

\ead{s.a.zaidi@leeds.ac.uk}

\address{School of Electronics and Electrical Engineering,\\
The University of Leeds, Leeds LS2 9JT\vspace{-0.3in}
}

\cortext[cor1]{Corresponding author}
\begin{abstract}
In this paper, we present an overview of Nearest neighbor (NN) methods,
which are frequently employed for solving classification problems
using supervised learning. The article concisely introduces the theoretical
background, algorithmic, and implementation aspects along with the
key applications. From an application standpoint, this article explores
the challenges related to the 5G and beyond wireless networks which
can be solved using NN classification techniques. 
\end{abstract}

\end{frontmatter}{}

\section{Introduction}

\subsection{Motivation}

Nearest neighbor (NN) classification belongs to a class of supervised
Machine Learning (ML) algorithms. The NN classification adopts NN
search techniques to solve the classification problem. The task of
classification involves the assignment of a class label to a given
query/sample data while utilising a Training dataset which has contains
several classified samples. Often sample points exist in a certain
metric space equipped with a well-defined distance function. The NN
search techniques find the sample in training data which has the least
distance to the unclassified query. This sample point is termed as
NN of the query. Subsequently, the class label associated with the
NN can be allocated to the unclassified query. The NN classification
algorithm is: i) simple to implement; ii) does not require knowledge
of the joint distribution of class and sample data; iii) performs
well in most problems especially for low-dimensional data-sets. Despite
the popularity of NN classification approaches there is a lack of
concise overview which provides a theoretical background while highlighting
algorithmic and implementation aspect. Even more importantly, while
classification problems are frequently encountered in the design and
analysis of modern wireless networks, a concise overview of the important
challenges and how they can be tackled using NN classification is
missing. To this end, this article is geared towards providing a comprehensive
overview of NN methods and their application for design and analysis
of 5G and Beyond Wireless Networks.

\subsection{Contributions \& Organisation}

The contribution of this article is two folds:
\begin{enumerate}
\item First, we provide an overview (see Section II) of the theoretical
and algorithmic framework for solving NN Search and Classification
problems. We also highlight how these techniques can be implemented
in practice. As an example, a Jupyter Notebook\citep{raza2020} is
provided as a supplement to this article. The notebook provides a
reference implementation on real data-set.
\item Second, we highlight (see Section III) key emerging scenarios related
to 5G and beyond wireless networks which present a certain classification
challenge. A comprehensive overview providing the context of the problem
and how some studies have cast them into thNN framework is provided.
Moreover, some of the emerging networking scenarios which present
similar challenges and yet remain unexplored are briefly mentioned.
\end{enumerate}

\section{Theoretical Framework}

In this section, we provide a concise overview of theoretical foundations
which characterise the NN classification techniques. We review mathematical
preliminaries and then explore relevant algorithms that are employed
to solve NN search (NNS) classification problems. 

\subsection{NN Search \& Classification}

\begin{mdframed}[backgroundcolor=nBlue!50,roundcorner=6pt,frametitlealignment=\centering, frametitle={NN Search Problem }]
Definition 1: Let $S$ be a set of objects and $d:S\times S\rightarrow\mathbb{R}$
be associated distance metric on $S$ metric space. Let $s_{i},s_{j},s_{k}\in S$,
then the function $d$ satisfies following three properties:

\noindent \ding{182} Positive property: $d\left(s_{i},s_{j}\right)>0$
for $s_{i}\neq s_{j}$;

\noindent \ding{183} Symmetric property: $d\left(s_{i},s_{j}\right)=d\left(s_{j},s_{i}\right)$;
and

\noindent \ding{184} Triangle inequality: $d\left(s_{i},s_{j}\right)\leq d\left(s_{i},s_{k}\right)+d\left(s_{j},s_{k}\right).$

\noindent Let $V\subseteq S$ be a certain subset of $S$ of size
$n$, then the nearest neighbor searching (NNS) problem is to build
a data structure, so that for an input query point say $q\in S$ an
element $v\in V$ is found with $d(q,v_{i})\leq d(q,v_{j})$ for all
$i\neq j,v_{j}\in V$.

\end{mdframed} 

\begin{table*}[t]
\vspace{-1in}
\centering\begin{tikzpicture} \node (table) [inner sep=0pt] {%
\begin{tabular}{|>{\centering}m{0.2\textwidth}|>{\raggedright}m{0.7\textwidth}|}
\hline 
\rowcolor{dnBlue}\textcolor{white}{Proximity Problem} & \textcolor{white}{Description}\tabularnewline
\hline 
\medskip{}

Fixed Radius NNS

\includegraphics[scale=0.2]{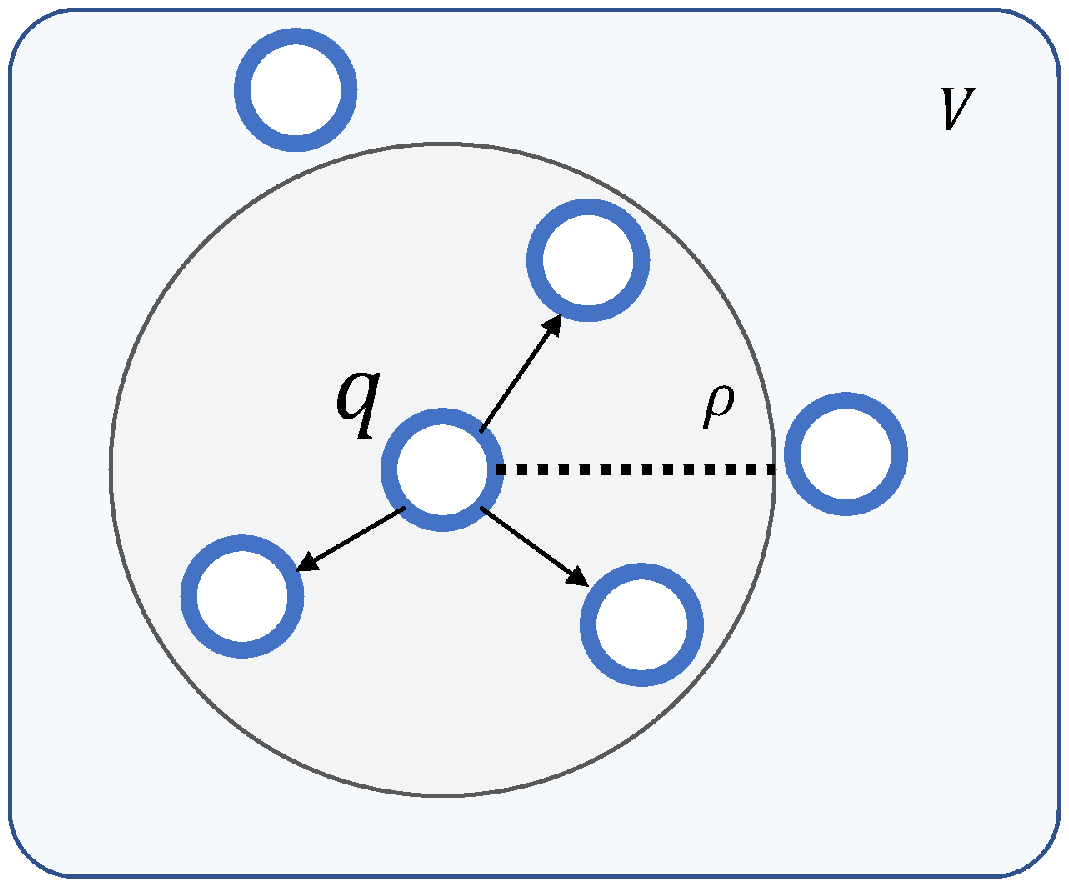} & The fixed radius nearest neighbor relaxes the exact NNS problem stated
in Definition 1, i.e. for a given query point $q\in S$, we are interested
in finding $R\subseteq V$ such that $d\left(q,r\right)\leq\rho$
for $r\in R$, with $\rho$ being the desired fixed radius.\tabularnewline
\hline 
\medskip{}
$k$-NNS

\includegraphics[scale=0.2]{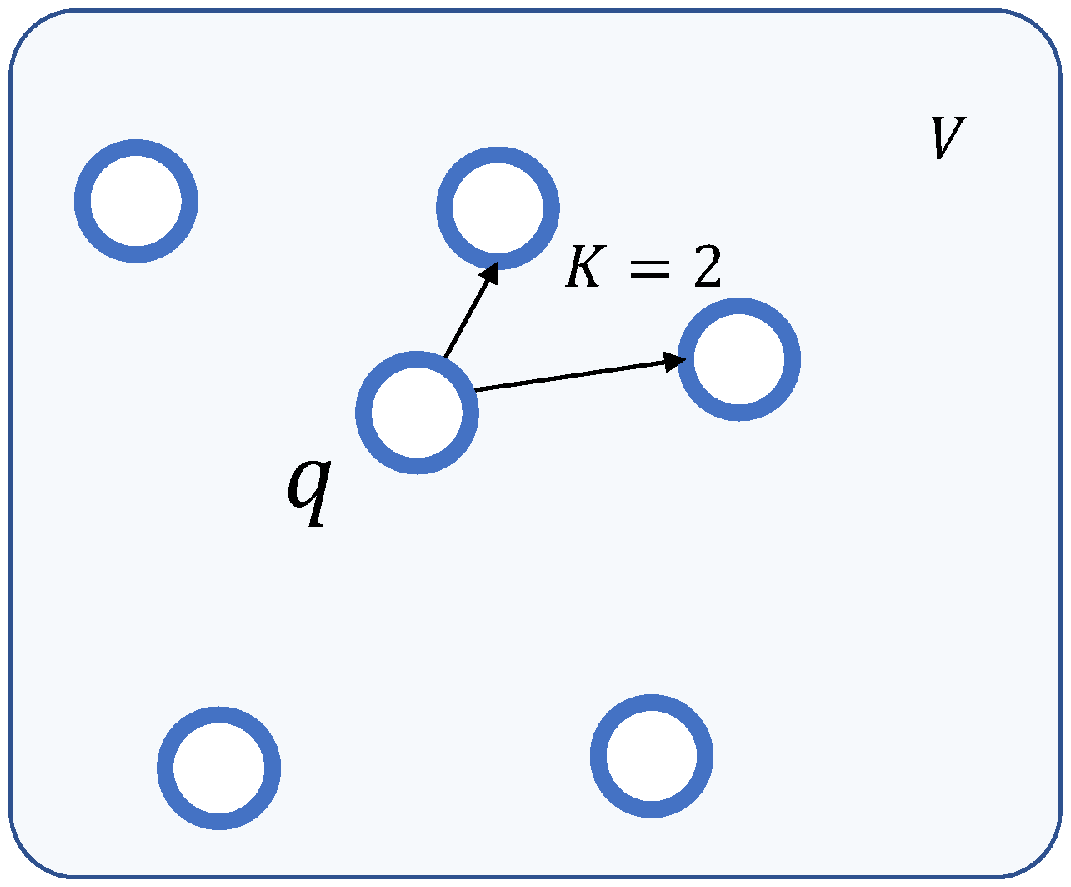} & The $k-$NNS problem tries to find $k\leq n$ nearest points in $V$
for a given query $q\in S$, i.e., the output for each $q$ is $R=\left\{ r_{i}:i=1..k\right\} $such
that $d\left(q,r_{1}\right)\leq d\left(q,r_{2}\right)\leq....\leq d\left(q,r_{k}\right)$.
In other words, if $q$ is the point in $\mathbb{R}^{d}$ then $k$-NNS
is geared to find closest $k$ points to $q$ which are contained
in set $V$.\tabularnewline
\hline 
\medskip{}
Minimum Spanning Tree

\includegraphics[bb=0bp 0bp 337bp 274bp,scale=0.2]{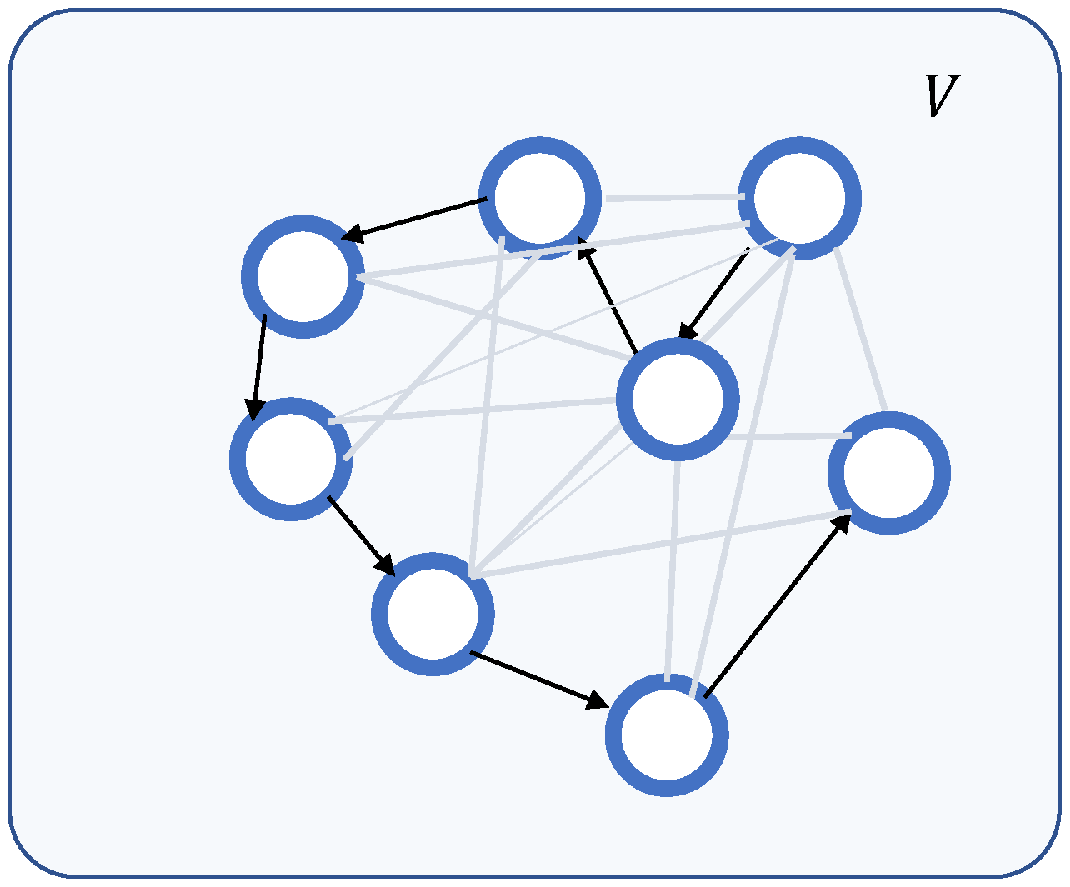} & A minimum spanning tree problem deals with finding a sub-set of edges
which connect all objects in $V$ without any cycles and with the
minimum possible total edge weight. In the context of proximity problems,
edge weight ($e_{ij}$) can simply be selected distance ($d\left(v_{i},v_{j}\right)$)
between every set of objects say $v_{i},v_{j}\in V$.\tabularnewline
\hline 
\medskip{}
Diameter Search

\includegraphics[scale=0.2]{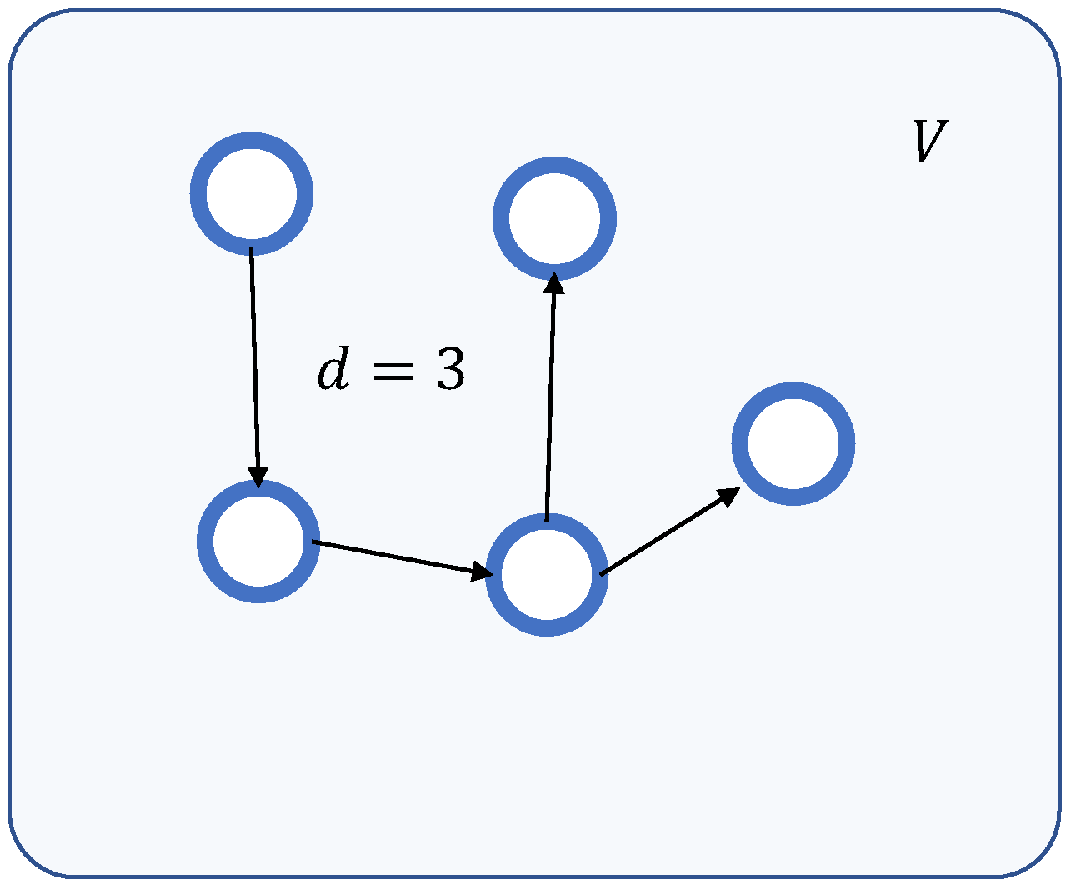} & The diameter problem is geared towards finding the two objects say
$v_{i},v_{j}\in V$ which are maximally distant from each other, i.e.
if distance between objects in the set $V$ is enumrated then the
maximum value for $d\left(v_{i},v_{j}\right)$ is the diameter.\tabularnewline
\hline 
\end{tabular}};\node[draw=dnBlue, inner sep=0pt, rounded corners=3pt, line width=2pt, fit=(table.north west) (table.north east) (table.south east) (table.south west)] {};\end{tikzpicture}\caption{Proximity Problems\label{tab:PProbs}}

\vspace{-0.2in}
\end{table*}
 The NNS problem is an example of so-called \textbf{proximity problems}
\citep{dickerson1996algorithms}, i.e. these are class of problems
whose definition requires association of a distance function with
the corresponding metric space. To provide a concrete overview of
the problem, we state a few well-known proximity problems in Table
\ref{tab:PProbs}. Notice that the list is not exhaustive and interested
readers are directed to \citep{dickerson1996algorithms,Clarkson2005NearestNeighborSA}
and references therein for detailed treatment. 

The NNS problem has received considerable attention from research
community whereby several techniques have been proposed to reduce
the neighbor search time. Typically $O(\log(n))$ can be achieved
by many proposed algorithms (See \citep{indyk1998approximate,muja2014scalable,datar2004locality,breuel2007note}
for details). The NNS is frequently employed to implement NN classification.
The classification problem involves assigning a label $\omega\in\mathcal{W=}\{1,2,...,M\}$
to $q\in S$ given a training set $T=\left\{ (s_{1},\omega_{1}),(s_{2},\omega_{2}),....,(s_{n},\omega_{n})\right\} $
where $s_{i}\in V\subseteq S$ and $\omega_{i}\in\mathcal{W}$.

\begin{mdframed}[backgroundcolor=nBlue!50,roundcorner=6pt,frametitlealignment=\centering, frametitle={NN Classification }]
Definition 2: The nearest neighbor classifier performs class assignment
for an unclassified sample point ($q\in S$) by assigning it the same
class label (say $\omega$) as those of the nearest set of previously
classified points. Generally, when such classification is performed
by inspecting a fixed number of enumerated distances say $k$, then
this classification is known as $k-$NN classification. $k-$NN is
one of the most frequently employed classifiers for supervised learning.
For $k=1$, the NN classification assigns $\omega=\omega_{i}$ where
$\omega_{i}$ is the label associated with $s_{i}$ which satisifies,
$d\left(q,s_{i}\right)\leq d\left(q,s_{j}\right)\forall s_{i},s_{j}\in V$.\end{mdframed} 

\subsection{Performance of NN Classification}

\begin{figure*}[t]
\centering{}\vspace{-1in}
\includegraphics[scale=0.5]{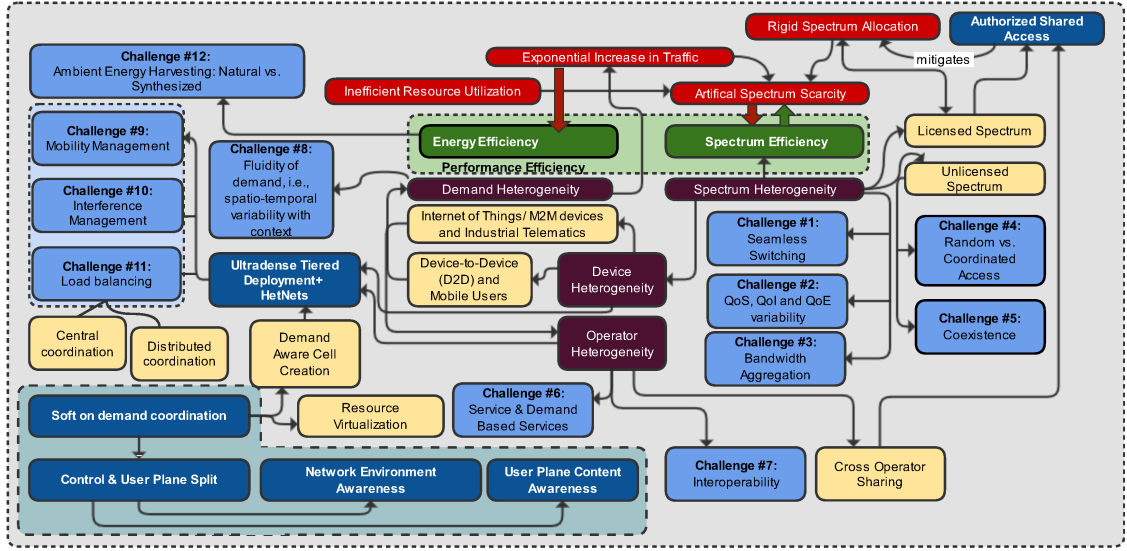}\caption{Performance Efficiency Metrics, Challenges, Solutions and Evolution
of 5G \& Beyond Networks.\label{fig:PEE}}
\end{figure*}
The accuracy of the classifier is measured in terms of error-rate
which is computed as an average of a loss function which encapsulates
the penalty of misclassification. Consider the test-set $T$ introduced
before, if the joint-distribution of $P(V,W)$ is completely known
than the Bayes classifier provides a minimum probability of error
$R^{*}$\footnote{The optimality of Bayes rule in providing a minimum probability of
error is non-trivial to proof. Interested readers are directed to
\citep{2018sltcastro,cover1967nearest}}. The NN classifier is a non-parametric and does not rely on the knowledge
of the joint distribution. For any number of classes, it has been
shown that the probability of error of the NN classifier is bounded
above by twice the Bayes probability of error\citep{2018sltcastro,cover1967nearest}.
In particular, when $\omega\in\mathcal{W=}\{1,2,...,M\}$ the probability
of error $R$ for the NN-classifier can be bounded as follows:
\begin{equation}
R^{*}\leq R\leq R^{*}\left(2-\nicefrac{(MR^{*})}{(M-1)}\right),
\end{equation}
these bounds are the tightest possible. 

\subsection{Algorithmic \& Implementation Aspects of NN Classification}

It is well known that the exact computation of NNS in higher dimensional
space is computationally expensive\citep{indyk1998approximate,10.1145/177424.177609}.
Specifically, when $S\subseteq\mathbb{R}^{d}$and the training set
has $n$ entries, the computational complexity for NNS is $O(dn)$.
Therefore, low complexity computation of $k-$NNS has been an open
area of research for several decades. There are a variety of solutions
which have been proposed so far. All these proposed approaches try
to compute approximate NN rather than the precise one. Naturally,
most of these algorithms have limitations, especially in higher dimensions.
Consider $\boldsymbol{q}\in S$ be the query or unclassified point,
let $\bar{\boldsymbol{s}}_{n}\in T$ be the NN of $\boldsymbol{q}$
in training set. The distance between$\bar{\boldsymbol{s}}_{n}$ \&
$\boldsymbol{q}$ be given by $d(\boldsymbol{q},\bar{\boldsymbol{s}}_{n})=\rho$,
then any point (say $\bar{\boldsymbol{q}}$) in $T$ which satisfies
the relationship $d\left(\bar{\boldsymbol{q}},\boldsymbol{q}\right)\leq(1+\epsilon)\rho$
is called $\epsilon$-NN of the query. There have been considerable
efforts from the research community to design algorithms that facilitate
the low-complexity computation of such $\epsilon$-NN search algorithms.
Details of such methods have been outlined by the Breuel in \citep{breuel2007note}.
The computational complexity of these approaches and distance metrics
that can be applied have been summarized in \citep{reza2014survey}
and are beyond the scope of this article. Nevertheless, in this section,
we would like to review one very popular approach which is available
in most modern ML frameworks (for instance SciKit Learn, TensorFlow,
MATLAB ML Toolbox, etc.).

\begin{figure}
\begin{centering}
\includegraphics[scale=0.5]{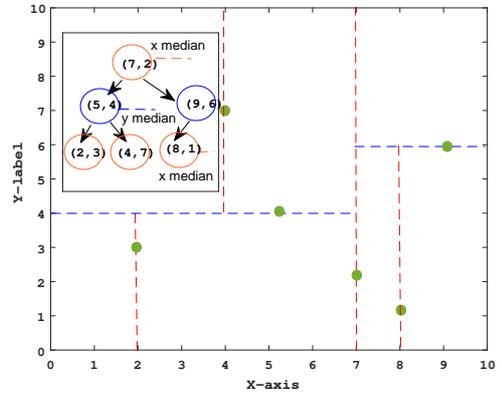}\caption{2-d Tree for a sample data $T=\{(7,2),(5,4),(9,6),(2,3),(4,7),(8,1)\}$\label{fig:2dtree}}
\par\end{centering}
\end{figure}

\noindent \textbf{$K-d$ Tree NN Search:} The $K-d$ tree algorithm
belongs to a class of algorithms that project higher dimensional data
onto lower dimensions thus speeding up the search. The core idea is
to split space into partitions that can be organised in form of trees,
enabling faster and localised search for the query within these partitions.
The $K-d$ tree algorithm is indeed generalisation of the binary tree
to $K$ dimensions. Each node in $K-d$ tree has one $K$ dimensional
key, two pointers (as its binary tree) which are either null or contain
a sub-tree and a discriminator between $0$ and $K-1$. Let $\textrm{Key}(P)=[\textrm{Key}_{o}(P),....,\textrm{Key}_{k}(P)]$
be the $K$-dimensional key associate with the node $P$, the two
pointers of $P$ can be denoted as $\textrm{Sub}_{1}(P)$ and $\textrm{Sub}_{2}(P)$.
Also, the discrimnator can be denoted by $\textrm{Disc}(P)$ for node
$P$. The tree is then constructed such that for any node $P$ in
a tree with $j=\textrm{Disc}(P)$, it is true that for any $Q\in\textrm{Sub}_{1}(P)$
it holds that $\textrm{Key}_{j}(P)>\textrm{Key}_{j}(Q)$. Similarly
for any node $Q\in\textrm{Sub}_{2}(P)$ it holds that $\textrm{Key}_{j}(Q)>\textrm{Key}_{j}(P)$.
All nodes on the same level in the tree have the same discriminator
(root having $0$ as discriminator) and it increases with the level
in the tree. This could be better understood with an example. The
construction can be better understood algorithmically as outlined
in Procedure \textsc{BuildKdTree} (see next page). Effectively the
algorithm splits the space across median values in each dimension.
A realisation for $2-d$ dataset is presented in the Fig. \ref{fig:2dtree}.
As is clear from the figure, the data points lie on the lines splitting
the plane. The tree representation is also shown in Fig. \ref{fig:2dtree}
where at each level split is performed around the median value of
$x-$ or $y-$axis in an alternative fashion. The splits of space
and organisation of the dataset in tree enable searching in smaller
sets speeding up the operation. $K-d$ tree can attain search complexity
of $O(\log(n))$. However, the building process has $O(n\log(n))$
time and $O(n)$ storage complexity. Further variants of $K-d$ include
the Ball tree method.

\begin{procedure}[t]
\caption{BuildKdTree(T,$depth$)}  
\SetAlgoLined 
\SetKwInOut{Input}{input}\SetKwInOut{Output}{output}
\Input{A set of points $T$ and the current depth $depth$} 
\Output{ The root of a $K-d$ Tree storing $T$}
initialization\; 
\eIf{$|T|==1$}{  
\Return $\textrm{Leaf}(T)$ \; 
\tcc{if there is only one point in $T$ return a Leaf containing that point}
}
{
\eIf{$depth\%2==0$ \tcp*{if $depth$ is even}}
{Split $T$ into two subsets $\textrm{Sub}_1$ and $\textrm{Sub}_2$ with a vertical line $l$ through median $x$-coordinate of the points.}
{Split $T$ into two subsets $\textrm{Sub}_1$ and $\textrm{Sub}_2$ with a horizontal line $l$ through median $y$-coordinate of the points.}
$v_\textrm{left}\leftarrow$  \textsc{BuildKdTree( $\textrm{Sub}_1$,$depth+1$)} \;
$v_\textrm{right}\leftarrow$  \textsc{BuildKdTree( $\textrm{Sub}_2$,$depth+1$)} \;
Create a node $\textrm{Node}(v)$ storing $l$, make $v_\textrm{left}$ the left child and $v_\textrm{right}$ the right child of $v$ \;
\Return $v$
} 
\end{procedure} 

\noindent \textbf{Implementation Issues: }Finally, we briefly want
to discuss the implementation of the $k-$NN algorithm. Let us consider
an indoor localisation scenario (see Section III for a detailed discussion).
We are often interested in identifying the propagation mode for the
received signal, i.e. whether the fixed anchor nodes (ANs) with known
locations and the tag (a device that needs to be localised) are in
Line-of-Sight(LoS) or Non-Line-of-Sight (NLoS). This binary classification
problem can easily be solved using the $k-$NN framework. The process
for accomplishing this is described in \textsc{Classify()} procedure.
A Jupyter notebook to accomplish this task in Python using SciKit
Learn on real-life data is provided with this article and can be downloaded
from \citep{raza2020}.

\begin{procedure}[t]
\caption{Classify(T)}  
\SetAlgoLined 
\SetKwInOut{Input}{input}\SetKwInOut{Output}{output}
\Input{A set of points $T={(x_1,y_1),(x_2,y_2),...,(x_n,y_n)}$ with labels and a test point $q$} 
\Output{ Class of $q$}
initialization\; 
Normalize $(x,y)$ \;
Split $T$ into Training $\alpha T$ and Testing Sets $(1-\alpha)T$ \;
Fit $k-$NN on $\alpha T$ \;
Use $k-$NN to validate performance on Test data $(1-\alpha)T$ \;
Use $k-$NN predict class label of q \;
\end{procedure} 

\section{Applications in Design \& Analysis of Emerging Communication Networks}

\subsection{Orchestration, Management and Allocation of Resources for 5G Network
Slicing}

5G \& beyond networks are envisioned to support three main service
classes: i) enhanced Mobile Broadband (eMBB); ii) massive machine
type communication (mMTC); and ultra-reliable low-latency communication
(URLLC) \citep{popovski20185g}. Each of these service classes is
suitable for a different set of verticals, for instance: i) eMBB services
can be employed for HD mobile video streaming; ii) mMTC services can
be employed for a sensor network solution for smart cities; and iii)
URLLC services can be employed for industrial automation. Each of
these service classes has a different service blueprint, i.e., different
throughput, latency, reliability, energy efficiency requirements.
Furthermore, each service type is geared to vary in terms of connection
densities, duty cycles, traffic profiles, and mobility dynamics. In
order, to address such demand, spectrum, and operator heterogeneity,
5G networks aims to adopt flexible and soft resource allocation. Fig
\ref{fig:PEE} graphically outlines performance efficiency challenges
addressed by 5G \& beyond networks and tools which are utilised to
mitigate the scarcity of resources. As demonstrated by the figure,
demand aware cell creation (see \citep{hashmi2018user}) and dynamic
allocation of resources is key for the cost-effective delivery of
such services. Dynamic and flexible resource allocation is enabled
through two key enablers, i.e., Network function virtualisation (NFV)
and Software Defined Networks (SDNs). Resources are managed, orchestrated,
and allocated in form of Network Slices (NSs) which are effective
logical partitions of physical compute, storage, and communication
resources\citep{zhang2017network}.

The orchestration of NS and its management while fulfilling the Quality-of-Service
(QoS) demands can be further enhanced by proactive demand forecasting\citep{8597639}.
In order to realise forecast aware slicer, the first and important
task is to classify the traffic and map it to pre-defined service
level agreement (SLA) blueprints (as articulated in 3GPP Specification
\citep{3gpptr}). The classification of traffic profiles can then
be employed by the forecasting engine which can request orchestration
of resources from the resource manager. In \citep{8597639,salhab2019machine}
and various other studies such architecture is proposed. The authors
have shown that $k-$NNs provide robust and low complexity classification
for the traffic profile. A variant of $k-$NN scheme is shown to achieve
around 95\% accuracy.

\subsection{Localisation \& Indoor Positioning}

Location-based services are core for enabling several 5G applications.
Indoor and outdoor localisation scenarios are characterised by different
propagation conditions and are of equal importance for different applications.
For instance, simultaneous localisation and mapping (SLAM) is a key
functionality required for the operation of industrial mobile robots
(MRs) and autonomous ground vehicles (AGVs). While outdoor localisation
can be enabled by current Global Navigation Satellite Systems (GNSS)
such as Global Positioning System (GPS), Galelio Positioning systems
etc. The current generation of GPS systems provide around $4.7\textrm{ m}$
and currently, industry giants (SpaceX, Lockheed Martin) are developing
a GPS III, which will be capable of providing $0.22-0.71$ m. The
positioning performance can be further improved by exploiting the
dense deployment of mmWave small cells. Indoor localisation albeit
is more challenging than outdoor due to harsh propagation conditions.
Indoor positioning has become more imminent due to the COVID-19 pandemic
and various solutions are currently being explored to enable contact
tracing.

The rapid proliferation of WiFi networks and dense deployments in
the urban indoor environment has triggered a lot of interest in exploring
indoor localisation based on WiFi Received Signal Strength (RSS).
The RSS based localisation schemes can be further classified into
two categories: i) Range-based localisation method, i.e., a method
which simply estimated RSS and path-loss exponent and then solve the
path-loss equation to find the distance from the anchor; ii) RSS fingerprinting
based localisation, i.e., a method which requires training data which
can be exploited to provide localisation estimate when sample data
at the unknown location is provided. The RSS fingerprinting method
is more robust as RSS readings can experience large variations in
indoor environments leading to higher positioning errors. 

\begin{mdframed}[backgroundcolor=nBlue!50,roundcorner=6pt,frametitlealignment=\centering, frametitle={RSS Fingerprinting as NN Classification }]The
RSS based fingerprinting can be easily caste into the NN classification
framework. Consider that, a user equipment (UE) device at a certain
location needs to be localised, then it is possible to build a vector
$\boldsymbol{x}=\{x_{1},x_{2},...,x_{N}\}$ where $\boldsymbol{x}\in\mathbb{R}^{N}$
is the vector containing RSS readings from $N$ access points (APS)
within the range of UE. If a training set $T$ exists such that $T=\left\{ (\boldsymbol{x}_{1},\boldsymbol{\omega}_{1}),....,(\boldsymbol{x}_{1},\boldsymbol{\omega}_{M})\right\} $
where $\boldsymbol{\omega}_{i}\in\mathbb{R}^{2}$ is the vector containing
coordinates and $\boldsymbol{x}_{i}\in\mathbb{R}^{N}$ be observed
RSS vector at this coordinates, then by performing NNS for the given
$\boldsymbol{x}$, we can find the class label $\bar{\boldsymbol{\omega}}_{n}$
associated with the NN $\bar{\boldsymbol{x}}_{n}\in T$ and therefore
can estimate the location of the AP. Of course, this requires $T$
to be large enough with fine spatial resolution. \end{mdframed} 

The NN algorithm can be further improved by selecting $k-$NNs and
selecting a maximally occurring label. This has indeed been explored
in \citep{xie2016improved,OH201891}. Further, improvement is possible
by considering APs with antenna arrays and utilising the channel state
information (CSI) instead of RSS. This has been recently explored
in \citep{sobehy2020csi}. Similar approaches can be adopted for finger-printing
RSS from Bluetooth enabled devices. 

Ultra-wideband (UWB) technology provides a very accurate localisation
with up to $10$ cm accuracy. However, such accuracy in an indoor
environment is only obtained when prior knowledge about the propagation
condition is present\citep{9205352}. For instance, classification
based on RSS value can serve as a first step to establish whether
there exists line-of-sight or non-line-of-sight propagation between
the anchor and the UE. This can then be used to refine time-difference-of-arrival
(TDoA) estimates. Naturally, the RSS classification can be cast into
$k-$NN framework.

\subsection{Beam Allocation Multiuser Massive MIMO}

\begin{figure}
\begin{centering}
\includegraphics[scale=0.4]{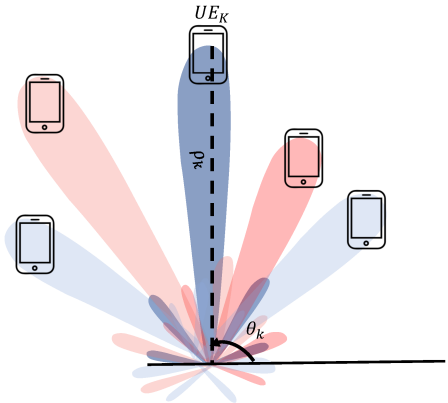}\caption{Illustration of Beam allocation in Multiuser Massive MIMO \citep{wang2018machine}.
The $K^{th}$ user is located at $\rho_{k}$ distance and at an angle
$\theta_{k}$.\label{fig:beam}}
\par\end{centering}
\vspace{-0.2in}

\end{figure}
Massive multiple-input multiple-output (MIMO) cellular networks are
envisioned as a key enabler for enhancing spectral efficiency in 5G
and beyond systems. Consider a beam allocation problem for the multiuser
Massive MIMO system as outlined in \citep{wang2018machine} and shown
in Fig \ref{fig:beam}. There are $K$ users each with a single receive
antenna. The base station (BS) is furnished with $N\gg K$ fixed beam
formed by Butler network with a linear array of $N$ identical and
isotropic radiating elements. The problem is how to allocate these
beams optimally to $K$ users to maximise the spectral efficiency.
Authors in \citep{wang2018machine} propose the use of cloud resources
to store training information. This training information is comprised
of beam allocation in response to a certain spatial configuration
of UEs. Then for any given set of $K$ users, $k-$NN classification
can be used to characterise the spatial pattern of users (encoded
in $(\rho_{k},\theta_{k})$). The $k-$NN classifier will mark the
pattern with the best class according to NNS and the class vector
can basically encode configuration pattern for beam allocation.

More recently, Reflecting Intelligent Surfaces (RIS) have been proposed
to enhance coverage and capacity in 6G wireless networks\citep{liu2020reconfigurable}.
The $k-$NN framework can also be adapted to optimise the configuration
of RIS elements. This however remains an open issue and has not been
extensively investigated.

\subsection{Sleeping Cell Anomaly Detection}

The sleeping cell problem (SCP) is a well-known problem in the Long
Term Evolution (LTE) cellular networks\citep{6963801,7145707}. The
SCP leads to cell outage and consequently lack of network services.
The SCP can manifest due to the Random Access Channel (RACH) failure\citep{7145707}.
Cell outage detection and identification of SCP conditions therefore
are one of the core features of big-data empowered self-organising
networks (SONs)\citep{6963801}. Under SCP cell outages cannot be
detected as the outage event manifests without triggering the alarms.
Consequently, no SON compensation function such as self-healing can
be deployed unless either multiple user complaints are received or
the outage is detected through drive testing. In such a scenario either
UEs can be configured to periodically report signal strength parameters
or neighboring cells can log such information. The measurement can
then be utilised by a $k-$NN classifier for anomaly detection, i.e.
labeling the readings as anomalous or normal. Anomaly detection can
lead to the rapid identification of SCP as shown in \citep{6963801}.
It has been shown by the authors that this approach yields $94\%$
detection accuracy. 

Anomaly detection is of importance in the context of network security
as well. Often intrusion is identified by observing anomalous traffic.
$k-$NN classifiers have been extensively used for such functionality.
Interested readers are directed to \citep{liao2002use} and references
therein.

\subsection{Energy Saving for Smart Devices}

Mobile devices have increasingly become a popular channel for digital
engagement. With ever-growing applications (Apps), energy consumption
and battery life-time are becoming core issues for manufacturers.
It is possible to exploit contextual information by building a device
and application profile to predict optimal configurations for the
wireless connectivity and location interfaces on the mobile device.
Such a technique can provide $24\%$ energy savings on average as
shown in \citep{6570479}. The authors demonstrate that the classification
and profiling can be attained by the $k-$NN classifier with $90\%$
accuracy. Similar techniques can be employed for the Internet-of-Things
(IoT) devices. Since most of the IoT devices are connected to the
cloud through the gateway, $k-$NN classification can be offloaded
to the cloud. This way optimisation of the configurations for energy
savings can be attained without incurring significant energy penalty
for obtaining such configuration on devices.

\section{Conclusion}

In this paper, we presented a brief overview of nearest neighbor methods.
We provided a concise review of the mathematical background and highlighted
the achievable performance for such methods when applied for solving
a classification problem. We briefly discussed algorithmic aspects
of such methods and highlighted key challenges in applying these techniques.
Based on the developed statistical and algorithmic framework, we review
some emerging applications where these classification techniques can
be employed.

\section*{Acknowledgment}

This work is supported by the EPSRC EP/S016813/1 and Royal Academy
of Engineering 122040 grants.

\vspace{-0.2in}
\bibliographystyle{elsarticle-num}
\addcontentsline{toc}{section}{\refname}\bibliography{biblo}

\end{document}